\documentclass[aps,prl,superscriptaddress, twocolumn, amsmath, amssymb]{revtex4}
\usepackage{graphicx}
\usepackage{amsmath,amssymb}
\usepackage{dcolumn}
\usepackage{mathrsfs}
\usepackage{physics}
\usepackage{float}
\usepackage{booktabs}
\usepackage{svg}
\usepackage{comment}
\usepackage{hyperref}
\usepackage{color}

\begin{document}
\title{A simple yet accurate stochastic approach to the quantum phase noise of nanolasers}
\date{\today}
	
\author{Matias Bundgaard-Nielsen}
\affiliation{Department of Electrical and Photonics Engineering, Technical University of Denmark, Building 343, 2800 Kongens Lyngby, Denmark}
\affiliation{NanoPhoton - Center for Nanophotonics, Technical University of Denmark, Building 343, 2800 Kongens Lyngby, Denmark}
\author{Marco Saldutti}
\affiliation{Department of Electrical and Photonics Engineering, Technical University of Denmark, Building 343, 2800 Kongens Lyngby, Denmark}
\affiliation{NanoPhoton - Center for Nanophotonics, Technical University of Denmark, Building 343, 2800 Kongens Lyngby, Denmark}
\author{Benjamin Falkenberg G\o tzsche}
\affiliation{Department of Electrical and Photonics Engineering, Technical University of Denmark, Building 343, 2800 Kongens Lyngby, Denmark}
\affiliation{NanoPhoton - Center for Nanophotonics, Technical University of Denmark, Building 343, 2800 Kongens Lyngby, Denmark}
\author{Emil Grovn}
\affiliation{Department of Electrical and Photonics Engineering, Technical University of Denmark, Building 343, 2800 Kongens Lyngby, Denmark}
\affiliation{NanoPhoton - Center for Nanophotonics, Technical University of Denmark, Building 343, 2800 Kongens Lyngby, Denmark}
\author{Jesper M\o rk}
\email[]{jesm@dtu.dk}
\affiliation{Department of Electrical and Photonics Engineering, Technical University of Denmark, Building 343, 2800 Kongens Lyngby, Denmark}
\affiliation{NanoPhoton - Center for Nanophotonics, Technical University of Denmark, Building 343, 2800 Kongens Lyngby, Denmark}

\begin{abstract}
Nanolasers operating at low power levels are strongly affected by intrinsic quantum noise, influencing both intensity fluctuations and laser coherence. Starting from semi-classical rate equations and making a simple hypothesis for the phase of the laser field, a simple stochastic model for the laser quantum noise is suggested. The model is shown to agree quantitatively with quantum master equations for microscopic lasers with a small number of emitters and with classical Langevin equations for macroscopic systems. In contrast, neither quantum master equations nor classical Langevin equations adequately address the mesoscopic regime. The stochastic approach is used to calculate the linewidth throughout the transition to lasing, where the linewidth changes from being dominated by the particle-like nature of photons below threshold to the wave-like nature above threshold, where it is strongly influenced by index fluctuations enhancing the linewidth. 
\end{abstract}
\maketitle

A nanolaser is also referred to as a cavity-QED laser since a large fraction of the spontaneous emission is emitted into the lasing mode, invalidating the phase transition picture of the onset of lasing \cite{Rice1994PhotonAnalogy}. Rather, the transition to lasing takes place over an extended interval of pump rates, where the quantum statistics of light change smoothly \cite{Rice1994PhotonAnalogy,Saldutti2024TheNanolasers}. Emerging nanolaser applications, such as on-chip communication \cite{Sun2015Single-chipLight}, programmable photonic integrated circuits \cite{Bogaerts2020ProgrammableCircuits}, sensing \cite{Ge2013ExternalBiosensor,Zhang2018ApplicationsCavities,Ma2019ApplicationsNanolasers}, as well as quantum technology \cite{Carolan2015UniversalOptics,Ma2019ApplicationsNanolasers,Wenzel2021SemiconductorRevisited} target ultra-low power operation and demand a detailed understanding of the quantum noise, which may be at a detrimentally high level \cite{Mork2020SqueezingConfinement}. While laser quantum noise can, in principle, be computed via master equation approaches \cite{Mu1992One-atomLasers,delValle2011RegimesExcitation,Gies2017StrongLasing,Bundgaard-Nielsen2023StochasticNanolaser}, such methods become unfeasible for more than a few emitters. Quantum Monte Carlo trajectory methods \cite{Molmer93} extend the range of feasibility by evolving a state vector instead of a density matrix but still suffer exponential scaling. Tensor network methods and matrix product states \cite{Strathearn2018,tensor_knorr_2020} reduce the Hilbert space by truncating the eigenvalues of a decomposed state vector, yet remain highly complex and have not yet been extended to nanolasers. Cumulant expansion methods \cite{Kubo1962GeneralizedCumulantExpansion,Plankensteiner2022quantumcumulantsjl,Drechsler2022RevisitingSiegertRelation} truncate higher-order quantum correlations in larger systems and are very promising but require a careful choice of truncation order, and have not been used to compute many-emitter laser linewidths so far. Meanwhile, semiclassical Langevin approaches \cite{Coldren1997}—successful for macroscopic semiconductor lasers—fail in the few-photon regime, leaving mesoscopic lasers unaddressed. Accurately describing and understanding the quantum noise in lasers is thus an important field that is still developing \cite{Kozlovskii2014Super-poissonianThreshold,Kreinberg2017EmissionCoupling, Gies2017StrongLasing, Mork2018RateEmitters, Andre2020, Takemura2021Low-Analogy, Protsenko2021QuantumNanolasers, Dimopoulos2022Electrically-DrivenThreshold, Yacomotti2023QuantumLimit, Carroll2021ThermalDevices, Vyshnevyy2022CommentDevices,Carroll2022CarrollReply:,Wang2020Superthermal-lightLasers,Bundgaard-Nielsen2023StochasticNanolaser,Pollnau2018PhaseAbsorption}. 

\begin{figure}
    \centering
    \includegraphics[width=0.93 \linewidth]{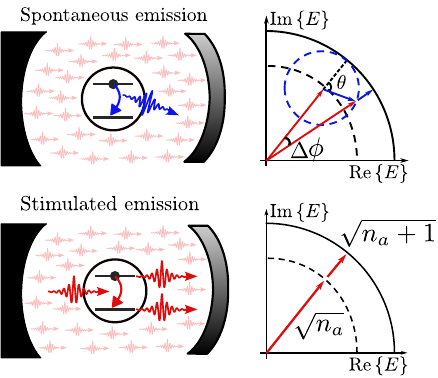}
    \caption{Illustration of the modeling of spontaneous and stimulated emission in the stochastic approach. A photon that is spontaneously emitted into the cavity mode induces a random phase shift, while a stimulated photon maintains the phase of the intracavity photon population. In both cases, the total number of photons is changed by one. }
    \label{fig:lw}
\end{figure}

It was previously shown that a stochastic approach based on classical laser rate equations yields a surprisingly accurate account of the intensity quantum noise of a single-emitter laser \cite{Bundgaard-Nielsen2023StochasticNanolaser}, while the phase noise and thereby the spectrum of the laser light could not be computed. Here, we show that by making a simple assumption about the phase of the intra-cavity laser field, the spectrum and, thereby, the laser linewidth can be computed. The results agree with full quantum mechanical simulations for small systems where collective effects can be neglected and with the standard Langevin approach for large systems. Compared to \cite{Bundgaard-Nielsen2023StochasticNanolaser}, we also show the applicability of the model beyond the one-emitter laser, and we show that the linewidth enhancement due to index fluctuations \cite{Henry1982TheoryLasers}, which typically dominates the linewidth of macroscopic lasers, follows naturally in the stochastic approach. This demonstrates that additional sources of phase noise can be incorporated and suggests that the approach can be broadly applied to other physical systems. For instance, extending it to optomechanics \cite{Rodrigues2010AmplitudeNoiseSuppression,Lorch2014LaserTheoryForOptomechanics,Bemani2017SynchronizationNanomechanicalMembranes,Xiong2023PhononPhotonLasingDynamics}  would be straightforward. Additionally, it may also be applicable to superconducting qubits \cite{Sears2012PhotonShotNoiseDephasing,Huang2016NoiseInducedTransitions,Guarcello2020VoltageDropLevyNoise} and Bose-Einstein condensates\cite{Cockburn2009StochasticGPE,Pieczarka2024BECPhotonsVCSEL}, which share similarities with laser systems. Using the new model, we are able to compute the laser linewidth throughout the transition region, lending new insights into the physics governing the coherence of lasers.

\begin{figure*}
    \centering
    \includegraphics[width= \linewidth]{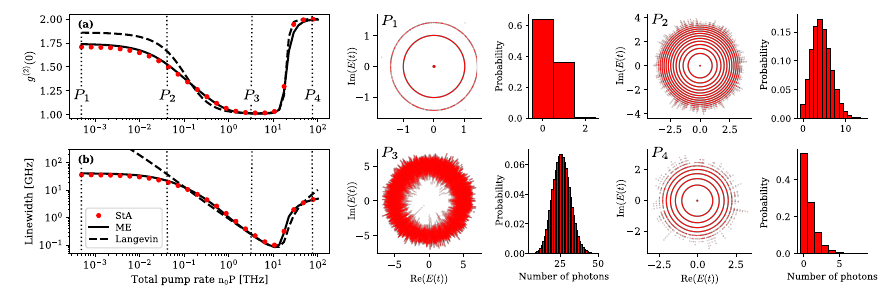}
    \caption{(a) second-order correlation function $g^{(2)}(0)$ and (b) linewidth as a function of pump-rate calculated for the StA, LA, and ME. 
    Insets $P_1$-$P_4$: The real and imaginary parts of the electric field, $E(t) = \sqrt{n(t)}\mathrm{exp}(i \phi(t))$ as well as the photon number distribution as produced by the StA for different pump rates indicated by dotted lines (a) and (b). Each point in the electric field plot represents the electric field at a single step in the simulation. Parameters are $g=0.1\ \mathrm{ps}^{-1}$, $\kappa = 0.04  \ \mathrm{ps}^{-1}$, $\gamma_A = 0.012 \ \mathrm{ps}^{-1}$, $\gamma_D=1 \ \mathrm{ps}^{-1}$, $n_0 = 5$ emitters, and $\alpha=0$.}
    \label{fig:phase_plot}
\end{figure*}

The starting point is the set of rate equations for the number of photons in the cavity mode, $n_a$, and the number of excited emitters, $n_e$, \cite{Mork2018RateEmitters}:
\begin{align}
      &\frac{d n_a}{dt} = \gamma_r (2n_e-n_0)n_a + \gamma_r n_e- \kappa n_a, \label{eq:na} \\
      &\frac{d n_e}{dt} = P (n_0-n_e) - \gamma_r (2n_e-n_0)n_a -\gamma_r n_e - \gamma_A n_e. \label{eq:ne}
\end{align}
Here, $\kappa$ is the cavity decay rate, $\gamma_A$ is the sum of the non-radiative decay rates and the decay rate into non-lasing modes, $n_0$ is the total number of emitters, and $P$ is the pump rate per emitter, modeled as incoherent pumping \cite{Mu1992One-atomLasers, Moelbjerg2013DynamicalEmitters}. We also introduced an emitter-cavity coupling rate given by $\gamma_r= 4g^2/(P + \kappa + \gamma_D + \gamma_A)$, with $g$ being the light-matter coupling rate and $\gamma_D$ the pure dephasing rate arising from electron-electron and electron-phonon scattering \cite{Strauf2011SingleNanolaser}. Note that $P$, $\gamma_D$, and $\gamma_A$ denote rates per emitter. These rate equations can be derived from quantum master equations by making a mean-field approximation and assuming that the material polarization can be adiabatically eliminated \cite{Strauf2011SingleNanolaser, Lorke2013TheoryTheory, Moelbjerg2013DynamicalEmitters}. This holds true when the dephasing of the system is large compared to the light-matter coupling rate $g$, i.e., $g<P+\kappa+\gamma_D +\gamma_A$, see Supplemental Material for a detailed discussion \cite{seeSL}.

The standard approach for including quantum noise in the mean-field equations \eqref{eq:na}-\eqref{eq:ne} is to add white noise Langevin forces $F_{a}$ and $F_{e}$ to the right-hand-side of the equations. In solving the subsequent stochastic differential equations, a small-signal analysis is often performed to obtain analytical results \cite{Coldren1997}. The inherent assumption is here that the noise terms $F_a$ and $F_e$ are small compared to the mean values $\bar{n}_a$ and $\bar{n}_e$, which is questionable when considering a laser with few emitters or a laser operating in a regime with few photons in the cavity mode. It was thus explicitly shown that the Langevin approach does not correctly predict the strong anti-bunching statistics of a single-emitter laser operating well below the threshold \cite{Bundgaard-Nielsen2023StochasticNanolaser}. Numerical simulations of the Langevin equations are also prone to numerical instabilities and have convergence issues \cite{Lippi2018NumericalPitfalls}. 

An alternative approach, where eqs.~\eqref{eq:na} and \eqref{eq:ne} are interpreted as describing a birth-death process \cite{Rice1994PhotonAnalogy} for integer-valued populations of photons and excited emitters and represented in the time domain as a stochastic process \cite{Puccioni2015StochasticLasing,Mork2018RateEmitters,Mork2020SqueezingConfinement,Andre2020} has been shown to be successful in predicting intensity noise in single-emitter nanolasers \cite{Bundgaard-Nielsen2023StochasticNanolaser}. The source of noise is both the stochastic evolution and the quantization of the variables, preventing the system from ever reaching "perfect equilibrium," leading to fluctuations around equilibrium values approximately given by the deterministic mean-field equations. 

Based on this success, we here hypothesize an extension to the stochastic approach (StA) that allows us to describe the phase of the electric field. We expand the analysis of \cite{Bundgaard-Nielsen2023StochasticNanolaser} to consider several emitters and, as we shall see, find accurate predictions of the linewidth of a nanolaser over a large range of parameter values. 
 
The main assumption is that we can describe the electric field of the cavity as:
\begin{equation}
    E(t) = \sqrt{n_a(t)}\mathrm{e}^{i \phi(t)} \mathrm{e}^{i \omega t},
\end{equation}
where $\phi$ is an additional stochastic variable describing the phase of the electric field. This allows calculating the emission spectrum, $S(\omega) = \int_0^\infty d \tau E(t+\tau) E^*(t) \mathrm{e}^{i \omega \tau}$. It is the evolution of the phase and eventual decoherence as the delay $\tau$ increases that gives the laser its finite linewidth above threshold. In describing the phase evolution, we adopt the following simple physical picture, first presented in refs. \cite{Henry1982TheoryLasers,Henry1986PhaseLasers}, which was shown to agree with the Langevin approach for large photon numbers. Photons emitted via stimulated emission are in phase with the already present photons. Spontaneously emitted photons, on the other hand, have a random phase, thus perturbing the phase and, over time, leading to a random drift. In Fig.~\ref{fig:lw}, we illustrate the implementation of this picture in the StA, where we show the real and imaginary parts of the electric field during stimulated and spontaneous emission events. For a stimulated emission event, the phase is unchanged, and the normalized electric field amplitude merely increases from $\sqrt{n_a}$ to $\sqrt{n_a+1}$. For a spontaneous emission event, we update the phase in the following way: Draw a random angle $\theta$ and take a unit-length step in that direction corresponding to a change of one photon to determine the new direction of the electric field, and thus the change of phase. 
Finally, since $n_a$ is restricted to integer values, the resulting electric field amplitude is projected onto the manifold associated with an increment of one in the photon number. In other words, the field amplitude transitions from $\sqrt{n_a}$ to $\sqrt{n_a +1}$ in a manner analogous to a stimulated emission event. We note that prior arguments have been made for fundamentally different phase models \cite{Zhao2016PhaseAdvance,Pollnau2018PhaseAbsorption} compared to the one proposed here, though without comparison to the prediction of full quantum simulations. The success of the present model corroborates our physical interpretation of the phase noise due to spontaneous emission. The algorithm behind the phase model is described in detail in the Supplemental Material \cite{seeSL}, and an open-source implementation is available at \cite{github}.

We compare the stochastic simulations to quantum master equation (ME) simulations of the Jaynes-Cummings model with $n_0$ identical emitters \cite{Loffler1997SpectralLaser,Mu1992One-atomLasers,Moelbjerg2013DynamicalEmitters,Bundgaard-Nielsen2023StochasticNanolaser}. Details are given in the Supplemental Material \cite{seeSL}. The computational effort for the ME simulations scales exponentially with the number of emitters, and we limit ourselves to consider at maximum $n_0=7$ emitters with the ME. To perform these calculations, we use QuantumOptics.jl in Julia \cite{Kramer2018QuantumOptics.jl:Systems} together with an Nvidia A100 Graphical Processing Unit (GPU). The GPU routines used are from ref.~\cite{Besard2019EffectiveGPUs}. 

In the following, we analyze a nanolaser consisting of $n_0=5$ emitters. We base our parameters on quantum emitters embedded in photonic crystal cavities and consider a light-matter coupling of $g=0.1\ \mathrm{ps}^{-1}$ \cite{Nomura2010LaserSystem}, a cavity decay rate $\kappa = 0.04  \ \mathrm{ps}^{-1}$ \cite{Dimopoulos2022Electrically-DrivenThreshold}, a non-radiative emitter rate of $\gamma_A = 0.012 \ \mathrm{ps}^{-1}$ and a pure dephasing rate of $\gamma_D=1 \ \mathrm{ps}^{-1}$. With these parameters, the adiabatic elimination of the polarization is well justified, and vacuum Rabi oscillations or other collective effects such as superradiance, which the rate equations cannot describe, are not important. 

Fig.~\ref{fig:phase_plot}(a) shows the intensity-correlation function $g^{(2)}(0)$ as a function of the total pump rate $n_0 P$ as calculated by the StA, ME, and LA. It is worth noticing the accuracy of the StA, whereas the LA deviates from the ME below threshold. This has already been observed before for a single-emitter laser \cite{Bundgaard-Nielsen2023StochasticNanolaser}, but we here demonstrate the accuracy for many emitters. From the $g^{(2)}(0)$-function, we also identify four different pump rates of interest, denoted, respectively, by $P_1$ to $P_4$. We investigate these points further in the insets of Fig.~\ref{fig:phase_plot} by showing the real and imaginary parts of the electric field, with each point representing a single timestep in the stochastic simulation. The corresponding photon number distributions are also shown.

The effect of spontaneous emission is evident at all four pump rates, with the temporal diffusion of the phase leading to circles in the phase space of the complex electrical field.

For $P=P_1$, we are below the lasing threshold, and the electric field is mostly in the zero photon manifold (middle dot). This is also clear from the photon number distribution. Occasional spontaneous emission events, however, lead to the electric field occupying the one- or two-photon manifolds. Here, the phases of subsequent photons are entirely uncorrelated, and the linewidth is dominated by the lifetime of the photons. The particle-like nature of the photons thus dominates.

For $P=P_2$, the laser is close to the lasing threshold, and the average photon population is higher. We see signatures of stimulated emission, evident from the "flares" extending from the circle formed by the electric field.

For $P=P_3$, we have reached lasing. In this case, one finds $g^{(2)}=1$, and the zero photon manifold is never occupied. Instead, we have a Poissonian photon distribution with an average of $\approx 25$ photons. The 
finite laser linewidth is now determined by the temporally diffusing phase rather than the lifetime of photons. It is now the wave-like nature of the photons that dominates.

Finally, for $P=P_4$, the pump-dephasing quenches the radiative rate and, consequently, the photon population, leading to the reoccupation of the zero manifold. Worth noting are the long "flares" extending from the main rings in the electric field, each representing a series of stimulated emission events characteristic of thermal or bunched light. This is further supported by $g^{(2)}(0)=2$.

In Fig.~\ref{fig:phase_plot}(b), we compare the pump-dependence of the linewidth obtained by the StA, ME, and LA (for details, see Supplemental Material \cite{seeSL}). The StA agrees very well with the ME. In deriving the linewidth from the LA, it was assumed that phase noise is the only contribution to the linewidth \cite{Henry1986PhaseLasers,Coldren1997,Wenzel2021SemiconductorRevisited}. This leads to a deviation for small pump values as intensity noise dominates in this regime. This highlights a general problem with the LA: Due to the complexity of solving the stochastic differential equations resulting from adding the Langevin forces, it is not possible to derive a valid expression from below threshold to above threshold. This is less of a problem in macroscopic lasers since the transition to lasing is very abrupt and well above the threshold the small-signal analysis of the LA correctly predicts the linewidth. For nanolasers, however, the transition to lasing is smooth with no unique and well-defined threshold \cite{Saldutti2024TheNanolasers}. The LA is thus not an adequate description since it is not obvious when it can be assumed to be valid. In contrast, the StA agrees with full quantum mechanical simulations both below, during, and above the lasing transition. 

In the Supplemental Material \cite{seeSL}, we compare the StA with the Schawlow-Townes linewidth expression \cite{Henry1986PhaseLasers}, extended with the usual corrections \cite{Cerjan2015QuantitativeLinewidth},  and find good agreement well below and well above threshold but no expression is available in the transition region. We chose here, along with \cite{Bjork1992OnLasers}, to show the analytical LA expression since it can be evaluated throughout the transition region and also predict the intensity noise correlations $g^{(2)}(0)$ in contrast to the Schawlow-Townes argument.

To further illustrate the versatility of the StA, we now consider a nanolaser with $n_0=1000$ emitters and $\gamma_A = 0.27 \ \mathrm{ps}^{-1}$, and $\kappa = 0.1 \mathrm{ps}^{-1}$, a scenario that currently cannot be computed using a full quantum mechanical description. Furthermore, we take into account that the refractive index depends on the number of excited carriers, an effect that is particularly important in semiconductor lasers \cite{Henry1982TheoryLasers}.
The effect is described by the linewidth enhancement factor (LEF) $\alpha = d n_r/d n_i$, which is the ratio of the change in the real part $n_r$ of the refractive index to the change in the imaginary part $n_i$ due to a variation in the number of excited emitters.
Assuming we are in a rotating frame with the cavity frequency, the phase now evolves according to \cite{Coldren1997}:
\begin{equation}
    \frac{d \phi}{d t} = \frac{\alpha}{2} \frac{d G}{d n_e} n_e = \alpha \gamma_r n_e, \label{eq:alpha}
\end{equation}
where $G=\gamma_r (2 n_e - n_0)$ is the net gain. If the number of excited emitters $n_e$ is constant, eq.~\eqref{eq:alpha} just results in a constant frequency shift. However, due to spontaneous emission, $n_e$ fluctuates, leading to frequency jitter and, thus, linewidth broadening. This effect is readily included in the StA by adding a second contribution to the phase equal to $\gamma_r n_e \alpha \Delta t$ at each timestep, where $\Delta t$ is the length of that timestep. This also serves as an example of how to include additional noise sources in the present phase model. 

\begin{figure}
    \centering
    \includegraphics[width=1 \linewidth]{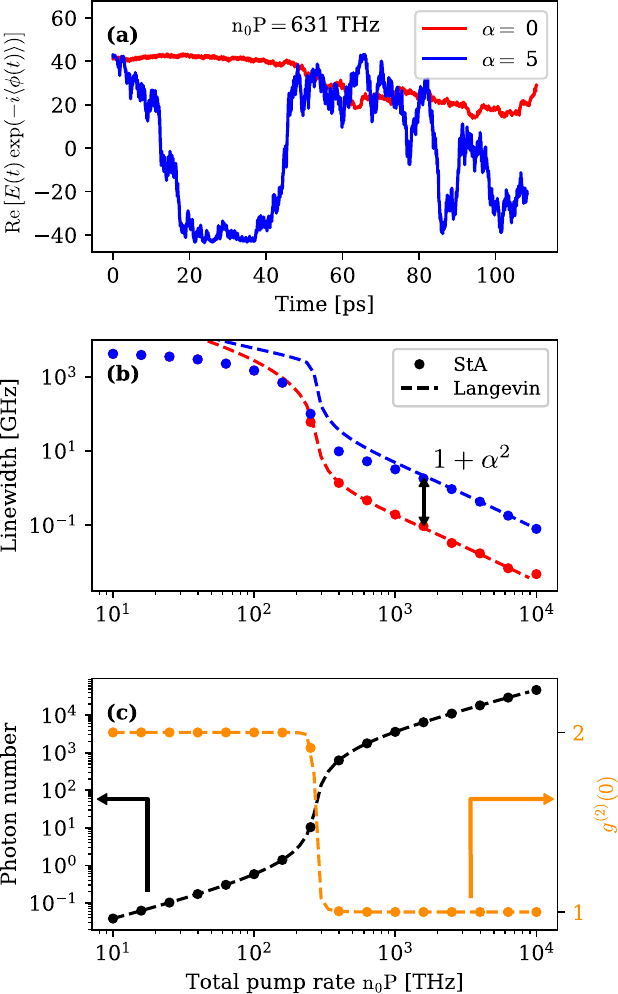}
    \caption{(a) The real part of the electric field $E(t)$ as a function of time for $P=1.0 \ \mathrm{THz}$ for $\alpha=0$ and $\alpha=5$. 
    (b) The linewidth as a function of the total pump rate $n_0 P$ for $\alpha=0$ and $\alpha=5$ as predicted by the LA and StA, respectively. (c) The photon number and intensity correlation function $g^{(2)}(0)$, as a function of the pump rate $n_0 P$, as predicted by the StA and LA. 
    Parameters are the same as in Fig.~\ref{fig:phase_plot} except $n_0=1000$, $\gamma_A = 0.27 \ \mathrm{ps}^{-1}$, and $\kappa = 2 \mathrm{ps}^{-1}$.}
    \label{fig:lw_alpha}
\end{figure}

In semiconductor lasers with bulk gain regions, the LEF is around $\alpha = 5-7$ \cite{Henry1982TheoryLasers,Henry1986PhaseLasers}, while lower-dimensional gain materials have smaller values, even approaching zero for some quantum dots \cite{Grillot2020PhysicsPhotonics}.  Fig.~\ref{fig:lw_alpha}(a) shows the evolution of the electric field $E(t)$ calculated using the StA for  $\alpha = 0$ and $\alpha=5$. We consider the regime above lasing with a total pump rate $n_0 P = 630 \ \mathrm{THz}$, where the effects of the LEF are pronounced. By subtracting the average diffusion of the phase $\expval{\phi(t)} =\alpha \gamma_r \bar{n}_e t$, we can compare the fluctuations for $\alpha=0$ and $\alpha=5$. It is clear that the phase of $E(t)$ fluctuates more for $\alpha=5$ than for $\alpha=0$, which leads to a larger linewidth. 

Fig.~\ref{fig:lw_alpha}(b) shows the computed linewidth versus total pump rate $n_0 P$. Below threshold, there is no significant dependence on  $\alpha$ since the linewidth is governed by the photon lifetime rather than the phase of the electric field. As the threshold is approached around \(n_0 P=270 \ \mathrm{THz}\), the linewidth for \(\alpha = 5\) begins to diverge from that for \(\alpha = 0\). For $\alpha = 0$, the linewidth decreases linearly with the photon number. In contrast, for $\alpha = 5$, the linewidth initially stagnates before decreasing linearly at higher pump rates. The linewidth for $\alpha = 5$ is $1 + \alpha^2 = 26$ times larger than that for \( \alpha = 0 \) above threshold, consistent with the correction to the Schawlow-Townes linewidth derived by C. Henry \cite{Henry1982TheoryLasers,Henry1986PhaseLasers} (see eq.~40 in Supplemental Material \cite{seeSL}). 

The linewidth predicted by the LA is also shown in Fig.~\ref{fig:lw_alpha}, and deviates from the StA in several ways. It diverges for small pump rates because it only includes phase noise, as was also seen in Fig.~\ref{fig:phase_plot}. More importantly, for $\alpha =5$, the neglect of intensity noise leads to deviations even well into the lasing regime (see also Eqs.~32-33 in Supplemental Material \cite{seeSL}). 

The transition to lasing in macroscopic semiconductor lasers has been analyzed before using the LA\cite{Agrawal1991IntensityLasers,Bjork1992OnLasers}. In those studies, it was predicted that the linewidth {\em increases} during the transition. However, the results of the StA show no such increase around the threshold. Instead, the linewidth gradually transitions to the modified Schawlow-Townes linewidth (see Supplemental Material eq. 40 \cite{seeSL}), highlighting the importance of including both intensity and phase noise. It is worth noting that we are considering $n_0=1000$ emitters, and the transition to lasing is well-defined and abrupt, as evidenced by the typical S-shaped input-output curve and the abrupt change of the quantum statistics of the light from thermal, $g^{(2)}(0)=2$, to Poissonian, $g^{(2)}(0)=1$, around $n_0 P\approx 270 \ \mathrm{THz}$, cf. Fig.~\ref{fig:lw_alpha}(c).

\begin{figure}
    \centering
    \includegraphics[width=\linewidth]{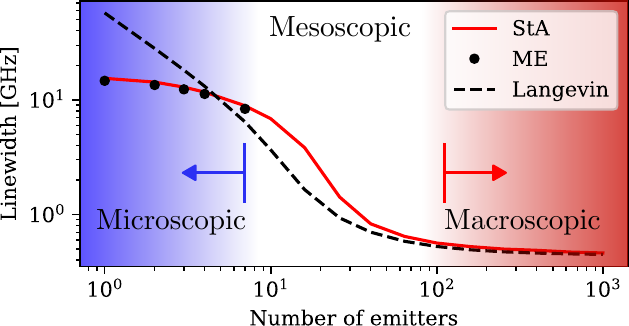}
    \caption{Linewidth versus number of emitters for $P=0.63 \ \mathrm{THz}$. Three regimes are shown: The microscopic regime, where full quantum mechanical calculations are feasible (ME); the mesoscopic regime; and the macroscopic regime, where conventional Langevin equations are valid. The stochastic approach works in all regimes. Parameters are the same as in Fig.~\ref{fig:phase_plot} }
    \label{fig:n0regimes}
\end{figure}

With our approach, we can address the mesoscopic regime of nanolasers with 10-100 emitters. Here, quantum descriptions are infeasible, and the validity of the Langevin approach is questionable. Fig.~\ref{fig:n0regimes} shows the linewidth as a function of the number of emitters for the three different approaches. The pump rate is fixed at twice the value necessary for inversion. Simulations using the ME approach are only computationally feasible for few emitters, denoted the microscopic regime, due to the large Hilbert space, while the Langevin approach is only valid for many emitters, denoted the macroscopic regime, due to the small-signal assumption. In contrast, the stochastic approach covers all regimes, including the important mesoscopic regime of $\sim 10-100$ emitters.

The stochastic model can be utilized to study complex mesoscopic laser systems. Such complex laser systems could include non-linear elements with the goal of reducing intensity noise or phase noise \cite{Sloan2023Driven-dissipativePhotonics,Nguyen2023IntenseFrequencies,Rivera2023CreatingContinuum,pontula2022strong}. With the stochastic approach one has access to the photon number distributions and dynamical phase portraits, providing a versatile tool that facilitates a better understanding of the quantum noise and dynamics of lasers, including the interplay between intensity noise and phase noise.

\begin{acknowledgments}
\section*{ACKNOWLEDGEMENTS}
This work was supported by the Danish National Research Foundation through NanoPhoton - Center for Nanophotonics, Grant No. DNRF147, and The European Research Council
(Grant No. 834410 FANO).
\end{acknowledgments}

\end{document}